\documentclass[12pt]{article}
\usepackage[english]{babel}
\usepackage{graphicx,subfigure}
\usepackage{amssymb,amsthm,amsmath,amsfonts}
\usepackage{float}
\usepackage[colorlinks=true,linkcolor=blue,citecolor=red, urlcolor=green]{hyperref} % per colorare citazioni, link, email ecc
\usepackage{cite}
\newtheorem{remark}{Remark}

\begin{document}

\title{An operatorial view of competition and cooperation in a network of economic agents}

\author{G.~Giunta${}^{*}$, M.~Gorgone${}^{**}$, F.~Oliveri${}^{**}$\\
\ \\
{\footnotesize ${}^{*}$ Fondazione MeSSinA, Parco Sociale di Forte Petrazza, 98151 Messina, Italy}\\
{\footnotesize g.giunta@fdcmessina.org}\\
{\footnotesize ${}^{**}$ MIFT Department, University of Messina,}\\
{\footnotesize Viale F. Stagno d'Alcontres 31, 98166 Messina, Italy}\\
{\footnotesize mgorgone@unime.it; foliveri@unime.it}
}

\date{Submitted.}

\maketitle

\begin{abstract}
A network of agents interacting both with competitive and/or cooperative mechanisms is modeled by using fermionic ladder operators. The time evolution of the network is assumed to be governed by a Hermitian time-independent Hamiltonian operator, and the mean values of the number operators are interpreted as a measure of the wealth status of the agents. Besides classical Heisenberg, we use the recently introduced $(H,\rho)$-induced dynamics approach to account for some actions able to provide a self-adjustment of the network according to its time evolution. Some numerical simulations are presented and discussed. Remarkably, we show that, in a network where cooperation may emerge, the average wealth of the agents is higher, and there is a very low level of inequality. 
\end{abstract}

$(H,\rho)$-induced dynamics; Network of economic agents; Competition and cooperation.

\section{Introduction}
Mathematical models whose ingredients are (fermionic or bosonic) ladder operators \cite{Roman,Merzbacher} have been proposed in the last two decades for describing several kinds of macroscopic systems \cite{bagbook1,bagbook2,fffbook,QSC,qdm1,qdm2}.
In fact, many recent contributions in the area of quantum-like modeling outside physics provided successful in different contexts: models of stock markets 
\cite{Bagarello-stock1,Bagarello-stock2,Bagarello-stock3,Bagarello-stock4}, 
social science and decision-making processes \cite{QSC,qdm1,qdm2,qdm3,qdm4}, political systems   
\cite{pol1,all1,all2,PKEH2016,BG17,DSO_turncoat2017,DSGO_turncoat2017,DSGO_opinion2017},  information spreading in a network \cite{BGO-information}. 

In an operatorial model of a macroscopic system $\mathcal{S}$, the unknowns  are operators acting on a Hilbert space $\mathbb{H}$ that can be finite- or infinite-dimensional (depending on the choice of using the fermionic rather than the bosonic representation). What we need is to identify the observables of  $\mathcal{S}$, \emph{i.e.}, the self-adjoint operators relevant for the description of the system itself, and 
compute the mean values of such operators evaluated on the state corresponding to an assigned initial condition; so doing we obtain some real valued functions that can be phenomenologically associated to some macroscopic quantities. 

The time evolution of such a system is found by introducing a self-adjoint 
time-independent Hamiltonian operator $\mathcal{H}$ embedding the various interactions occurring 
among the actors of the system. In particular, to take the computational complexity low, in the 
following we will limit ourselves to a quadratic Hamiltonian. This has as an immediate consequence on the time evolution which is at most quasiperiodic. Nevertheless, we may add another ingredient using the recent approach of $(\mathcal{H},\rho)$-induced dynamics \cite{BDSGO-PhysicaA,fffbook} by assuming  that at fixed times some checks on the variation of the mean values of the observables in a small time interval produce a change in some of the parameters entering the Hamiltonian, without modifying its functional form. This approach, that provided successful in many situations \cite{DSO_turncoat2017,DSGO_turncoat2017,DSGO_opinion2017}, allows 
us to describe a sort of discrete self-adaptation of the model depending on the evolution of the 
state of the system. We stress that in this strategy the rule $\rho$ is not introduced as a mere mathematical expedient, but  is somehow physically justified as it determines a possible change of attitudes and interactions of the agents of the system $\mathcal{S}$.

In this paper, we implement and investigate numerically a model for a system $\mathcal{S}$ 
made of a finite number of agents whose mutual interactions can be thought of as competitive or cooperative. Each agent ${A}_j$ is associated to an annihilation ($a_j$) and a  creation ($a_j^\dagger$)  fermionic operator, and the mean value associated to the number operator $\widehat{n}_j=a_j^\dagger a_j$ is interpreted as a measure of the wealth of  ${A}_j$. In \cite{coop-ijtp}, a first model where a system made of competing and/or cooperating agents has been described. Anyway, the approach and the results hereafter presented are rather different.

The structure of the paper is the following. 
In Section~\ref{sec:model}, a short review of the number representation for fermions is given, and the 
time-independent self-adjoint  quadratic Hamiltonian operator incorporating the interactions is analyzed. 
Subsection~\ref{sec:rule} briefly describes the $(\mathcal{H},\rho)$-induced dynamics approach; 
the  physical meaning of the \emph{rules} we consider is also clarified. 
In Section~\ref{sec:numericalsimulations}, we present and discuss various numerical simulations. Finally,
Section~\ref{sec:conclusions} contains some concluding remarks. 

\section{The operatorial model}
\label{sec:model}
In order to implement an operatorial model for describing the evolution of a system $\mathcal{S}$ made of $N$ 
interacting agents $A_j$, $j=1,\ldots,N$, an annihilation ($a_j$), a creation ($a^\dagger_j$), and an 
occupation number ($\widehat n_j=a^\dagger_j a_j$) fermionic operator are associated to each agent. 
These fermionic operators satisfy the \emph{canonical anticommutation relations} (CAR)
\begin{equation}
\label{CAR}
\{a_i, a_j\}=0,\quad \{a_i^\dagger, a_j^\dagger\}=0,\quad
\{a_i, a_j^\dagger\}=\delta_{i,j}\mathbb{I},
\end{equation}
$i,j=1,\ldots, N$, $\mathbb{I}$ being the identity operator, and $\{u,v\}=uv+vu$ the 
anticommutator between $u$ and $v$. 
These operators act on a  Hilbert space $\mathbb{H}$ linearly spanned by the orthonormal set of vectors
\begin{equation}
\label{fermion_vectors}
\varphi_{n_1,n_2.\ldots,n_N}=
(a_1^\dagger)^{n_1}(a_2^\dagger)^{n_2}\cdots(a_N^\dagger)^{n_N}\varphi_{0,0,\ldots,0}, 
\end{equation} 
generated by acting on the \emph{vacuum} $\varphi_{0,0,\ldots,0}$ (\emph{i.e.}, an eigenvector of 
all the annihilation operators) with the operators 
$(a_i^\dagger)^{n_i}$, $n_i=0,1$ for $i=1,\ldots,N$; therefore, it is 
$\dim(\mathbb{H})=2^N$.  
The vector $\varphi_{n_1,n_2,\ldots,n_N}$ means that to the $i$th agent it is initially assigned 
a mean value equal to $n_i$ ($i=1,\ldots,N$). We have 
\begin{equation} 
\widehat n_i\varphi_{n_1,n_2,\ldots,n_N}=n_i\varphi_{n_1,n_2,\ldots,n_N},\qquad i=1,\ldots,N. 
\end{equation}
The interpretation we give to the mean values $n_i$ ($i=1,\ldots,N$) is that of a measure of the 
wealth of the $i$th agent. 

Let us assume the dynamics of $\mathcal{S}$ to be governed by the Hermitian time-independent Hamiltonian
\begin{equation}
\mathcal{H}=\mathcal{H}_0+ \mathcal{H}_I,
\label{hamiltonian}
\end{equation}
where
\begin{equation}
\left\{
\begin{aligned}
\mathcal{H}_0 &= \sum_{j=1}^N\omega_j a_j^\dagger a_j,\\
\mathcal{H}_I&=
\sum_{1\le i<j \le N}\lambda_{i,j}\left(a_i\,a_j^\dagger+a_j\,a_i^\dagger\right)+\sum_{1\le i<j \le N}\mu_{i,j}\left(a_i^\dagger\,a_j^\dagger+a_j\,a_i\right),
\end{aligned}
\right.
\end{equation}
the constants $\omega_j$, $\lambda_{i,j}$ and $\mu_{i,j}$ being real positive quantities; we 
remark that in concrete applications not all parameters $\lambda_{i,j}$ and $\mu_{i,j}$ need to be different from zero. 

The contribution $\mathcal{H}_0$ is the free part of the Hamiltonian, and the parameters $\omega_{j}$ are 
somehow related to the inertia of the operators associated to the agents of 
$\mathcal{S}$: in fact, they can be thought of as a measure of the tendency of each degree of 
freedom to stay constant in time \cite{bagbook1,fffbook}. Thus, they can describe the attitudes of the agents according to their greater or lesser inclination to change.

On the contrary, $\mathcal{H}_I$ embeds the interactions among the agents. These split 
in two contributions:
\begin{itemize}
\item the term $\lambda_{i,j}\left(a_i\,a_j^\dagger+a_j\,a_i^\dagger\right)$ can be interpreted as 
a competitive  contribution, and the coefficient $\lambda_{i,j}$  gives a measure of the strength 
of the interaction between the agents $A_i$ and $A_j$;  in fact, the term 
$a_{i}a_{j}^\dagger$ destroys a \emph{particle} for the agent 
associated to $a_{i}$ and creates a \emph{particle} for the agent associated to $a_{j}$; the 
adjoint term $a_{j}a_{i}^\dagger$ swaps the roles of the two agents;
\item the term $\mu_{i,j}\left(a_i^\dagger\,a_j^\dagger+a_j\,a_i\right)$ can be interpreted as a 
cooperative contribution, and $\mu_{i,j}$ is a measure of the strength of this cooperation;
the term $a_i^\dagger a_j^\dagger$ creates a \emph{particle} for both agents, and the adjoint part 
destroys a \emph{particle} for both agents. 
\end{itemize}

Adopting the Heisenberg view for the dynamics, the time evolutions of the annihilation operators $a_j(t)$ are ruled by 
\begin{equation}
\frac{d a_j}{dt} =\textrm{i} \left[H,a_j\right] ,\qquad j=1,\ldots,N,
\end{equation}
$[\mathcal{H},a_j]=\mathcal{H}a_j-a_j\mathcal{H}$ being the commutator between $\mathcal{H}$ 
and $a_j$. Thus, we have a system of linear ordinary differential equations (whose unknowns are operators), namely 
\begin{equation}
\label{eq-annihilation}
\begin{aligned}
\frac{d a_j}{dt}&=\textrm{i}\left(-\omega_j a_j + 
\sum_{1\le \ell < j\le N}\left(\lambda_{\ell,j} a_\ell + \mu_{\ell,j} a_\ell^\dagger\right)\right.\\
&\left.+\sum_{1\le j< k \le N }\left(\lambda_{j,k} a_k -\mu_{j,k} a_k^\dagger \right)\right),
\end{aligned}
\end{equation}
that have to be solved with suitable initial conditions $a_j(0)=a_j^0$, $j=1,\ldots,N$. Because 
each operator $a_j$ is a square matrix of order $2^N$, in principle we have to solve a system of $N\cdot 4^N$ linear 
ordinary differential equations.

Nevertheless, because of the linearity, we can adopt a \emph{reduced} approach. Let us introduce a 
formal vector
$\mathbf{A}\equiv\left(a_1,\ldots,a_N,a_1^\dagger,\ldots, a_N^\dagger\right)^T$ 
(the superscript ${}^T$ stands for transposition) and the square matrix of order $2N$ 
\[
\Gamma= \left[
\begin{array}{cc}
\Gamma_0 & \Gamma_1 \\
-\Gamma_1 & -\Gamma_0 
\end{array}
\right],
\]
where the symmetric block $\Gamma_0$ and the antisymmetric block $\Gamma_1$ are
\[
\Gamma_0 = \left[
\begin{array}{ccccc}
-\omega_1 & \lambda_{1,2} & \cdots & \cdots & \lambda_{1,N} \\
\lambda_{1,2} & -\omega_2 & \lambda_{2,3} & \cdots & \lambda_{2,N}  \\
\cdots & \cdots & \cdots & \cdots & \cdots \\
\cdots & \cdots & \cdots & \cdots & \cdots \\
\lambda_{1,N}, & \lambda_{2,N} & \cdots & \lambda_{N-1,N} & -\omega_N
\end{array}
\right]
\]  
and
\[
\Gamma_1 = \left[
\begin{array}{ccccc}
0 & -\mu_{1,2} & \cdots & \cdots & -\mu_{1,N} \\
\mu_{1,2} & 0 & -\mu_{2,3} & \cdots & -\mu_{2,N}  \\
\cdots & \cdots & \cdots & \cdots & \cdots \\
\cdots & \cdots & \cdots & \cdots & \cdots \\
\mu_{1,N}, & \mu_{2,N} & \cdots & \mu_{N-1,N} & 0
\end{array}
\right],
\]  
respectively.

With these positions, equations~(\ref{eq-annihilation}), together with their adjoint version, 
write in the compact form
\[
\frac{d\mathbf{A}}{dt}=\textrm{i}\Gamma \mathbf{A},\qquad \mathbf{A}(0)=\mathbf{A}^0,
\]
whose solution formally is
\[
\mathbf{A}(t)= \mathcal{B}(t)\,\mathbf{A}^0, \qquad \mathcal{B}(t)=
\exp\left(\textrm{i}\Gamma t\right).
\] 
Now, let us define the vector 
\[
\Phi=\sqrt{n_1^0}\varphi_{1,0,\ldots,0}+\sqrt{n_2^0}\varphi_{0,1,\ldots,0}+\cdots+\sqrt{n_N^0}\varphi_{0,0,\ldots,1},
\]
where $(n_1^0,n_2^0, \ldots, n_N^0)$, such that $n_1^0+\cdots+n_N^0=1$ ($\Phi$ is unitary), represent the initial values of the mean values of the number 
operators associated to the agents of the system.  
If $B_{i,j}$ is the generic entry of matrix $\mathcal{B}(t)$, we have
\begin{equation}
\begin{aligned}
&a^\dagger_j(t)=\sum_{i=1}^{N}\left(B_{j+N,j}a_i^0+B_{j+N,i+N}{a_k^0}^\dagger\right),\\
&a_j(t)=\sum_{i=1}^{N}\left(B_{j,i}a_i^0+B_{j,i+N}{a_i^0}^\dagger\right),
\end{aligned}
\end{equation}
whereupon the formula
\begin{equation}
\label{observable}
n_j(t)=\left\langle\Phi,a^\dagger_j(t)a_j(t)\Phi\right\rangle,\qquad j=1,\ldots,N,
\end{equation}
using the canonical anticommutation relations (\ref{CAR}), provides
the mean values of the number operators at time $t$:
\begin{equation}
\begin{aligned}
n_{j}(t)&=\sum_{i=1}^{N}\Phi_{i}^2\sum_{\ell=1}^{N}B_{j,f(\ell,j)}B_{j+N,g(\ell,j)}\\
&+\sum_{i=1}^{N-1}\sum_{\ell=i+1}^{N}\Phi_i\Phi_\ell\left(B_{j,i}B_{j+N,\ell+N}+B_{j,j}B_{j+N,i+N}
\right.\\
&\left.\qquad-B_{j,i+N}B_{j+N,\ell}-B_{j,\ell+N}B_{j+N,i}\right),
\end{aligned}
\end{equation}
where
\[
f(\ell,j)=j+(1-\delta_{\ell, j})N,\qquad
g(\ell,j)=j+\delta_{\ell, j}N,
\]
$\delta_{\ell, j}$ being the Kronecker symbol.

In our modelization, the real functions given by (\ref{observable}) stand for the measures of the wealth status of the agents of the system. 

\subsection{$(\mathcal{H},\rho)$-induced dynamics}
\label{sec:rule}
The Heisenberg dynamics can be modified  by superposing periodically some specific \emph{rules} 
that introduce some effects on the evolution that cannot 
be embedded in the definition of the Hamiltonian, and that do not introduce technical or computational difficulties (see \cite{BDSGO-PhysicaA,fffbook}, and references therein). 

The  action of the rules modifies some of the values of the parameters involved in the Hamiltonian as a consequence of the evolution of the system. In such a way,  the model adjusts itself during the time evolution changing the attitudes of the agents, the strength and/or the nature of the interactions.

The  steps to be done adopting the $(\mathcal{H},\rho)$-induced dynamics are listed below (see \cite{BDSGO-PhysicaA,fffbook}, and references therein for further details):
\begin{enumerate} 
\item choose an initial condition;
\item divide the time interval $[0,T]$ in $n$ subintervals of length $\tau$;
\item let $k=1$, and consider a Hamiltonian operator $\mathcal{H}^{(k)}$;
\item using Heisenberg view,  compute, in the time interval $[(k-1)\tau,k\tau]$, the evolution of ladder operators, and consequently of the mean values of the number operators; 
\item according to the  variations in the time interval $[(k-1)\tau,k\tau]$ of these mean values,  modify some of the parameters involved in $\mathcal{H}^{(k)}$; 
\item a new Hamiltonian operator $\mathcal{H}^{(k+1)}$, having the same functional form as $\mathcal{H}^{(k)}$, but (in general) with different values of (some of) the involved parameters, is obtained; 
\item increment by 1 the value of $k$;
\item if $k<n$, go to step 4.
\end{enumerate}

Glueing the local evolutions in all the subintervals, we get the global evolution of the system: therefore, the latter is governed by a sequence of similar Hamiltonian operators, and the parameters entering the model result stepwise (in time) constant.

Next Section will be focused on a concrete example, and the evolution is studied by means of numerical simulations considering various situations.

\section{Numerical simulations}
\label{sec:numericalsimulations}
In this Section, we numerically study a system made of $N=100$ economic interacting agents. First of all, we have to fix the initial values of the parameters entering the Hamiltonian.  
The agents are randomly partitioned in three subgroups with different ranges of the initial inertia parameters, say 
\begin{itemize}
\item 25 agents have inertia parameters randomly chosen in the interval $[0.2,0.4]$; 
\item 50 agents have inertia parameters randomly chosen in the interval $[0.5,0.7]$; 
\item 25 agents have inertia parameters randomly chosen in the interval $[0.8,1.0]$. 
\end{itemize}
Then, we define a pseudo distance among the agents $A_i$ and $A_j$, namely
\begin{equation}
\label{distance}
d_{i,j}=\left|\omega_i-\omega_j\right|,
\end{equation}
that will be used to set the strength of the interactions. As already observed, the inertia parameters, in some sense, describe the attitudes of the agents, and the pseudo distance \eqref{distance} can be thought of as a measure of their more or less similarity.

As far as the interactions are concerned, the initial choice is detailed below:
\begin{itemize}
\item $N_{comp}$ agents (randomly chosen) form the competitive subgroup: among them we select randomly $2N_{comp}$ couples and fix the strength of their mutual competition as
\begin{equation}
\label{comp_pars}
\lambda_{i,j}=0.1(1+\tanh(3 d_{i,j})),
\end{equation}
so that the competition parameter between a couple of agents increases as the pseudo distance between their attitudes increases;

\item $N_{coop}$ agents (randomly chosen) form the cooperative subgroup: among them we select randomly $2N_{coop}$ couples and fix the strength of their mutual cooperation as
\begin{equation}
\label{coop_pars}
\mu_{i,j}=0.1(2-\tanh(3 d_{i,j})),
\end{equation} 
so that the cooperation parameter between a couple of agents increases as the pseudo distance between their attitudes decreases;

\item the remaining (if any) $N_{opp}=N-N_{comp}-N_{coop}$ agents form the opportunist subgroup: each member of this subgroup competes with a randomly chosen agent in the competitive subgroup and cooperates with a randomly chosen agent in the cooperative subgroup. The strength of competition and cooperation is assumed as above.
\end{itemize}

\begin{remark}
Of course, not all couples of agents are initially interacting. The rationale of taking the competition and cooperation parameters according to the relations \eqref{comp_pars} and \eqref{coop_pars} is the following: interacting agents with \emph{similar} attitudes, \emph{i.e.}, with close inertia parameters, privilege cooperation to competition, the opposite when the inertia parameters have a higher pseudo distance. We observe that competition and cooperation parameters, where not vanishing, belong to the interval $[0.1,0.2]$.
\end{remark}

Finally, the initial condition is such that all agents start with the same amount of wealth.

The distribution of wealth $n_i(t)$ among the agents can be analyzed by means of the Gini index \cite{Gini-index},
\[
G(t) = \frac{\sum_{i,j=1}^N|n_i(t)-n_j(t)|}{2N\sum_{i=1}^N n_i(t)},
\]
belonging to the interval $[0,1]$; it is a measure of statistical dispersion, and in our case gives a measure of the wealth inequality. A Gini index close to 0 corresponds to an almost uniform wealth distribution, whereas a value close to 1 to a wealth distribution with strong inequalities.

\begin{figure}
\centering
\subfigure[$N_{coop}=10$, $N_{opp}=50$]{\includegraphics[width=0.49\textwidth]{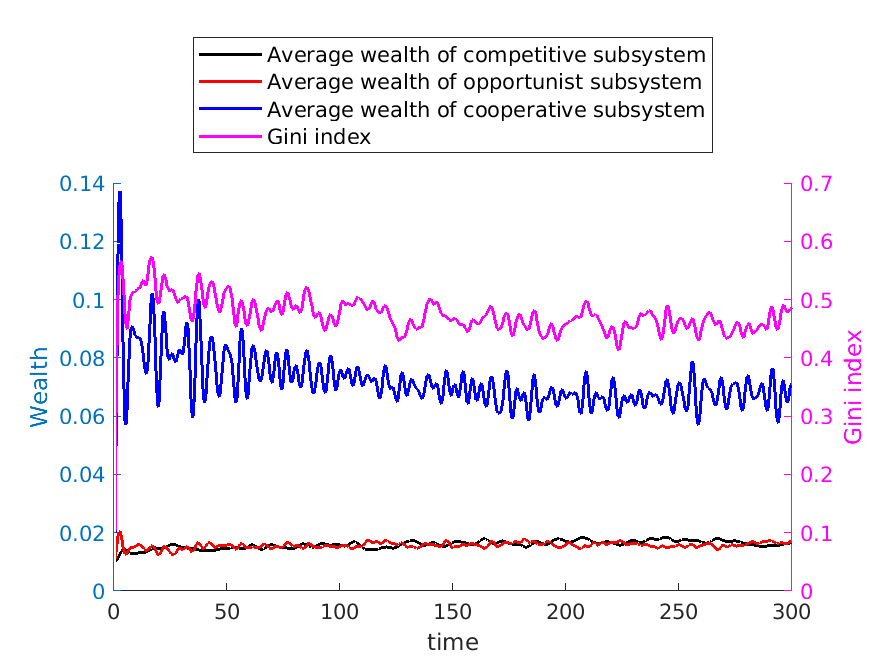}\label{fig1}}
\subfigure[$N_{coop}=20$, $N_{opp}=40$]{\includegraphics[width=0.49\textwidth]{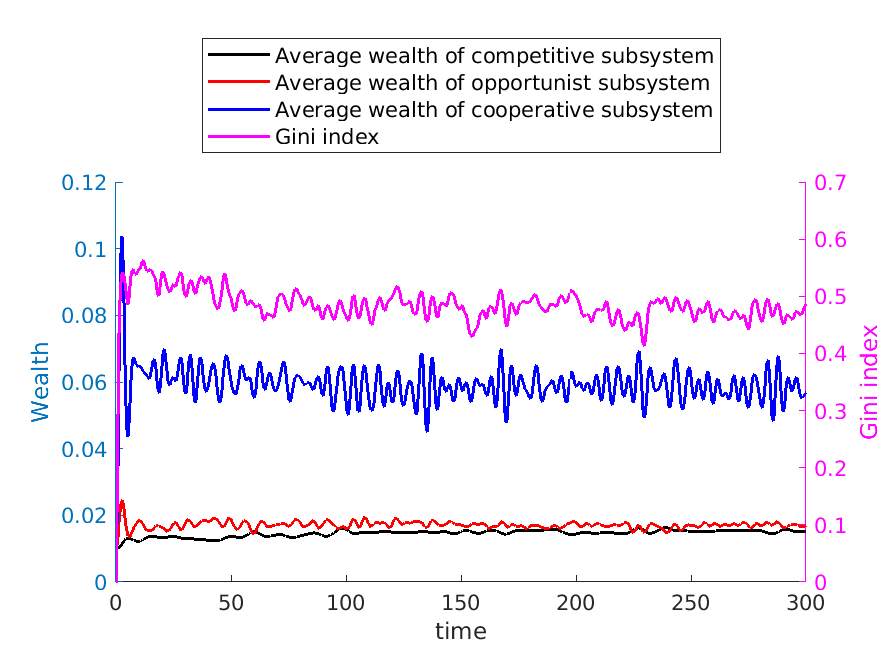}\label{fig2}}\\
\subfigure[$N_{coop}=30$, $N_{opp}=30$]{\includegraphics[width=0.49\textwidth]{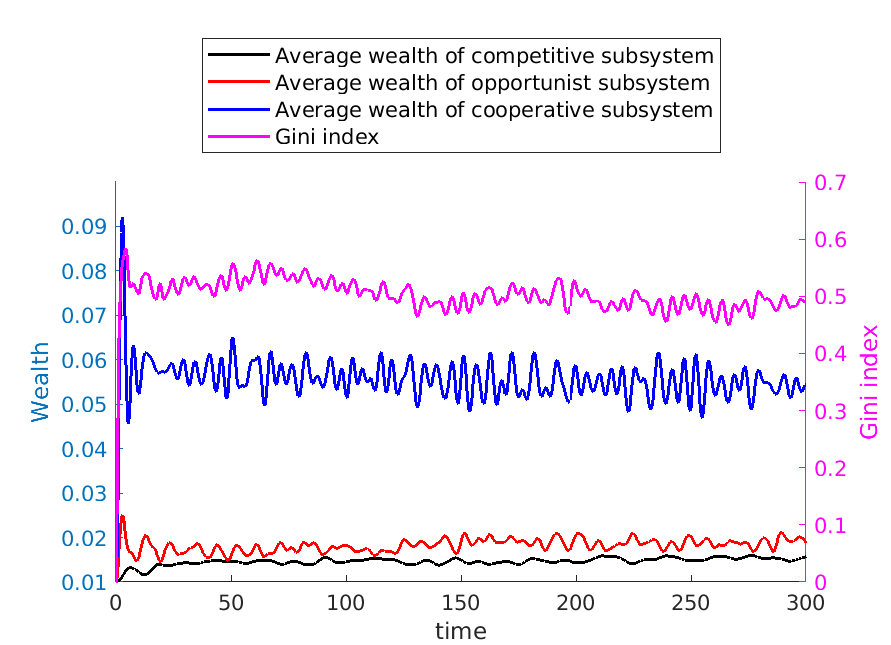}\label{fig3}}
\subfigure[$N_{coop}=40$, $N_{opp}=20$]{\includegraphics[width=0.49\textwidth]{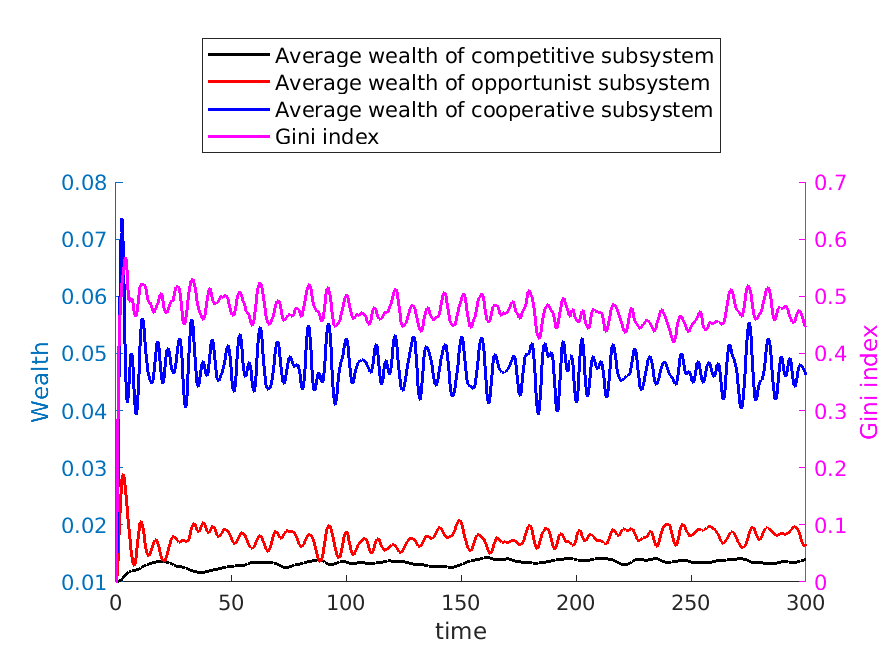}\label{fig4}}
\caption{Standard Heisenberg dynamics. Average wealth of the three subgroups of agents, and Gini index for the distribution of wealth among the agents of the system, with $N_{comp}=40$.  The scale on the left refers to wealth, that on the right to Gini index.}
\end{figure}

First of all, we use the classical Heisenberg dynamics, and the time evolution of the average amount of wealth for the three subgroups is displayed in Figures~\ref{fig1}-\ref{fig4}; in all the cases, it is $N_{comp}=40$, whereas the values for $N_{coop}$ is chosen equal to 10 (Figure~\ref{fig1}), 
20 (Figure~\ref{fig2}), 30 (Figure~\ref{fig3}), 40 (Figure~\ref{fig4}). In the same figures, the Gini index vs. time is also displayed. 

Whatever the size  of the cooperative subgroup is, we observe that the average wealth status of the cooperative subgroup is higher than that of the remaining subgroups. Anyway, looking at the Gini index, that initially is zero, we observe that it exhibits a trend with small oscillations around the value 0.5. This means that the evolution of the system generates inequality among the agents.

The situation can drastically change if we adopt the  $(\mathcal{H},\rho)$-induced  dynamics approach. In particular, we choose $\tau=1$ as the length of the time subinterval where the parameters entering the Hamiltonian are kept constant; at instants $k\tau$ ($k$ positive integer)  the parameters are modified as specified below.
Let us define
\begin{equation}
\begin{aligned}
&\delta_{j}^{(k)}=n_{j}(k\tau)-n_{j}((k-1)\tau), \qquad j=1,\ldots,N,\\
& \delta^{(k)}=\hbox{max}\left\{ \left\vert \delta_j^{(k)} \right\vert ,\;  j=1,\ldots,N\right\};
\end{aligned}
\end{equation}

We consider two different set of rules. In the first scenario, at times $k\tau$ we update the inertia parameters, say 
\begin{equation}
\label{rule_inertia}
\omega_{j} \, \rightarrow \, \omega_{j} \left( 1+\frac{\delta_{j}^{(k)}}{\delta^{(k)}} \right); 
\end{equation}
we observe that the agents (possibly) change their inertia but their interactions are not modified at all.
The relation \eqref{rule_inertia} indicates that the inertia parameter of the agent $A_j$ increases (decreases, respectively) when its wealth status increases (decreases, respectively) in the subinterval. The rationale of the rule is that an agent whose wealth status is increasing becomes more conservative and less susceptible to change. 

\begin{figure}
\centering
\subfigure[$N_{coop}=10$, $N_{opp}=50$]{\includegraphics[width=0.49\textwidth]{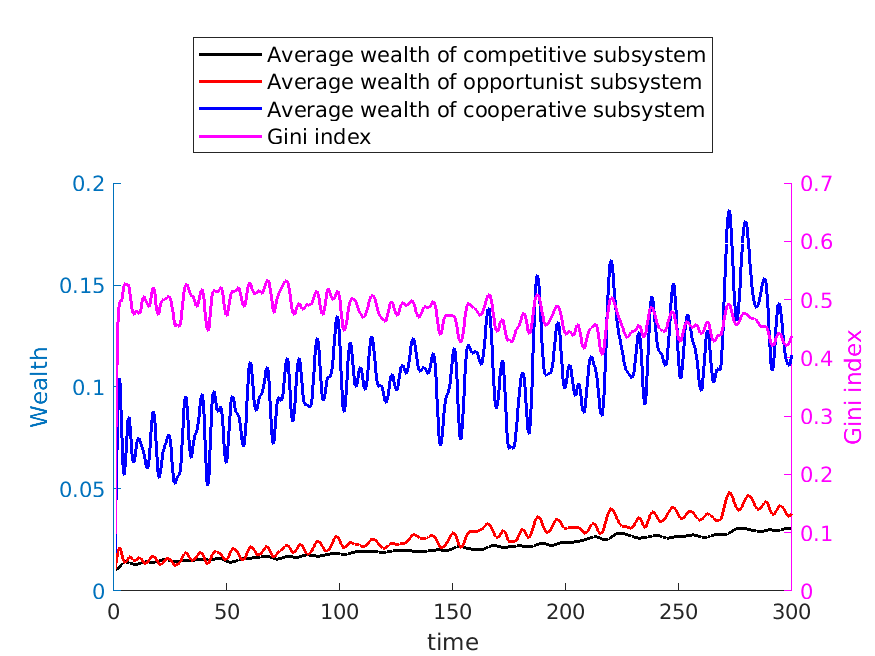}\label{fig5}}
\subfigure[$N_{coop}=20$, $N_{opp}=40$]{\includegraphics[width=0.49\textwidth]{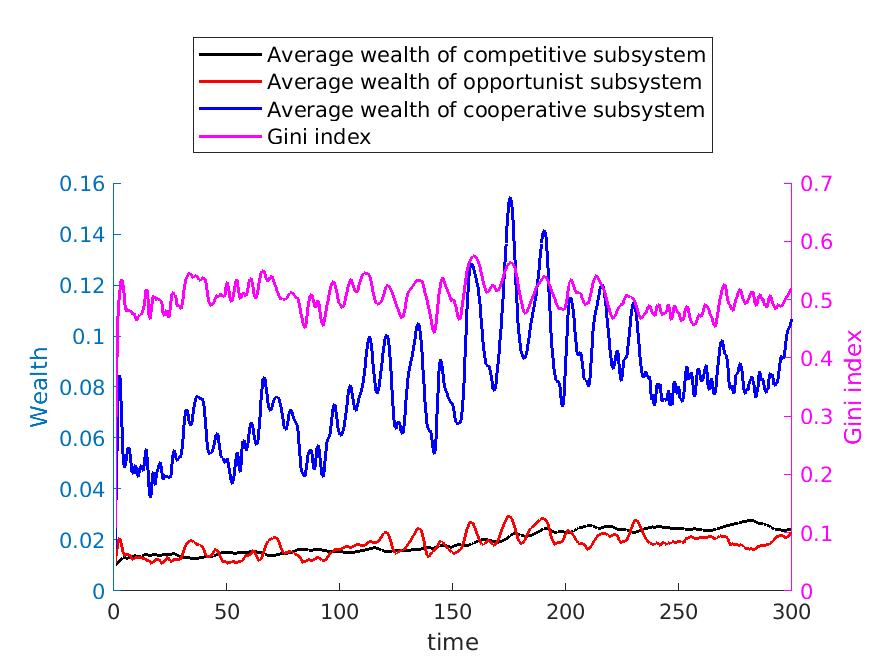}\label{fig6}}\\
\subfigure[$N_{coop}=30$, $N_{opp}=30$]{\includegraphics[width=0.49\textwidth]{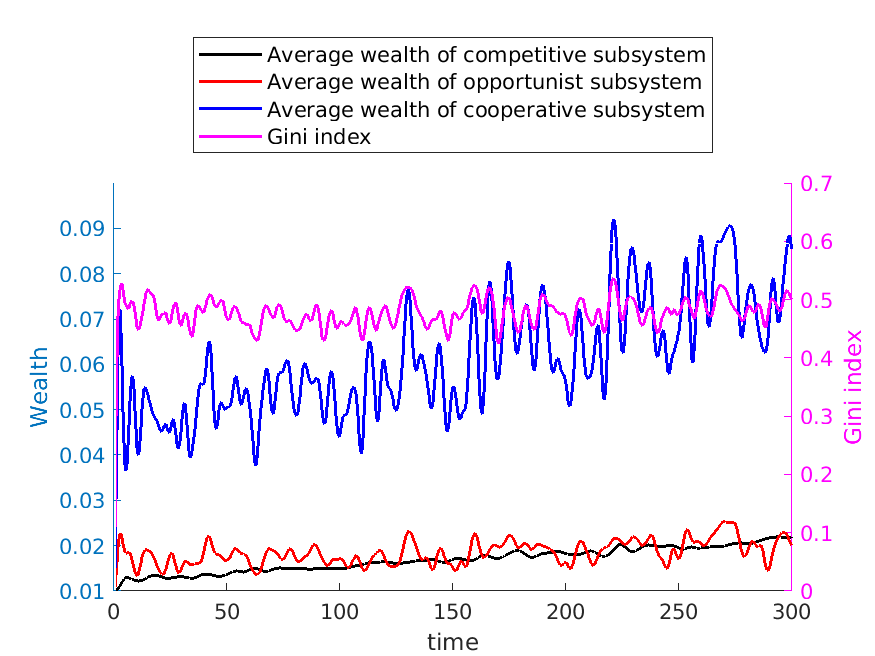}\label{fig7}}
\subfigure[$N_{coop}=40$, $N_{opp}=20$]{\includegraphics[width=0.49\textwidth]{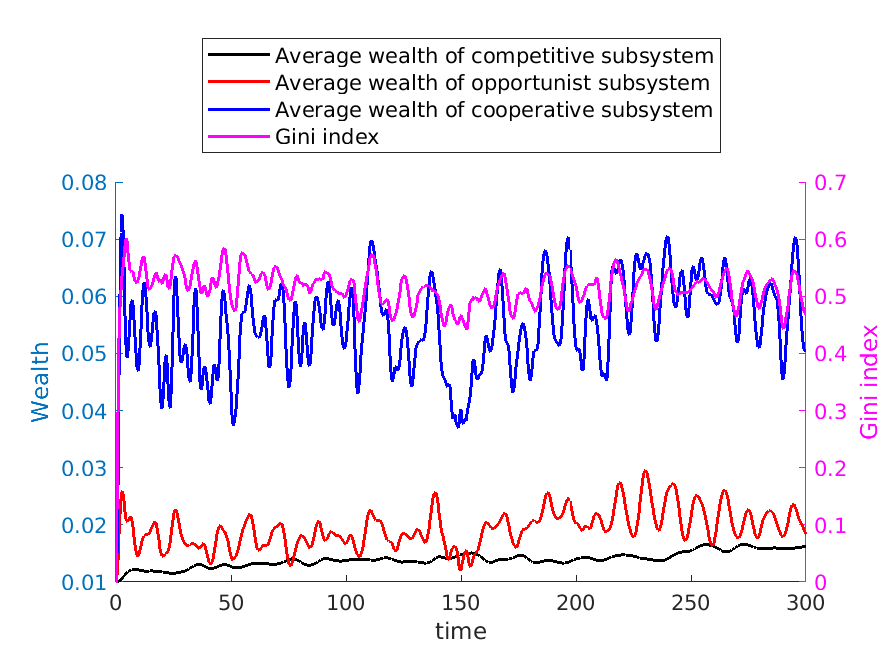}\label{fig8}}
\caption{$(\mathcal{H},\rho)$-induced dynamics with rules modifying only inertia parameters. Average wealth of the three subgroups of agents, and Gini index for the distribution of wealth among the agents of the system, with $N_{comp}=40$.  The scale on the left refers to wealth, that on the right to Gini index.}
\end{figure}

The results of the numerical simulations are displayed in Figures~\ref{fig5}-\ref{fig8}.

What can be observed is that, assuming the agents to change their attitudes, the cooperative subgroup has an average wealth still higher than the other subgroups. Moreover, competitive and opportunist subgroups have average wealth similar to those in the classical Heisenberg dynamics, whereas the evolution of the wealth status of cooperative subgroup is higher that that obtained without adopting the rule. Furthermore, we stress that the evolution seems to be not affected by the size of the cooperative subgroup.
 
In the second scenario, besides updating the inertia parameters using \eqref{rule_inertia}, we adopt the following rules for changing the interaction parameters:
\begin{equation}
\label{rule_interaction}
\begin{aligned}
&\hbox{if}\; \delta_i^{(k)}>0\; \hbox{and}\;\delta_j^{(k)}>0\;\hbox{then}\\
&\qquad \mu_{i,j}\mapsto
\hbox{min}\left(\mu_{i,j}+\delta_i^{(k)}+\delta_j^{(k)},\mu_{max}\right), \\
&\qquad \lambda_{i,j}\mapsto
\hbox{max}\left(\lambda_{i,j}-\delta_i^{(k)}-\delta_j^{(k)},\lambda_{min}\right),\\
&\hbox{else if}\;\delta_i^{(k)}<0\;\hbox{and}\; \delta_j^{(k)}<0\;\hbox{then}\\
&\qquad \mu_{i,j}\mapsto
\hbox{max}\left(\mu_{i,j}+\delta_i^{(k)}+\delta_j^{(k)},\mu_{min}\right), \\
&\qquad \lambda_{i,j}\mapsto 
\hbox{min}\left(\lambda_{i,j}-\delta_i^{(k)}-\delta_j^{(k)},\lambda_{max}\right),
\end{aligned}
\end{equation}
where $\lambda_{min}=\mu_{min}=0$ and $\lambda_{max}=\mu_{max}=0.2$.

The relations \eqref{rule_interaction}, which are able to change both the number of interactions in the system, as well as their nature and strength, is a sort of \emph{win-win} rule: in fact, a couple of agents whose variations of wealth status are positive for both increases the strength of cooperation, and decreases the strength of competition; the opposite change occurs if both agents undergo a negative variation of their wealth status. Moreover, this rule can produce the birth of new interactions, as well as the removal of some existing interactions.
	
\begin{figure}
\centering
\subfigure[$N_{coop}=10$, $N_{opp}=50$]{\includegraphics[width=0.49\textwidth]{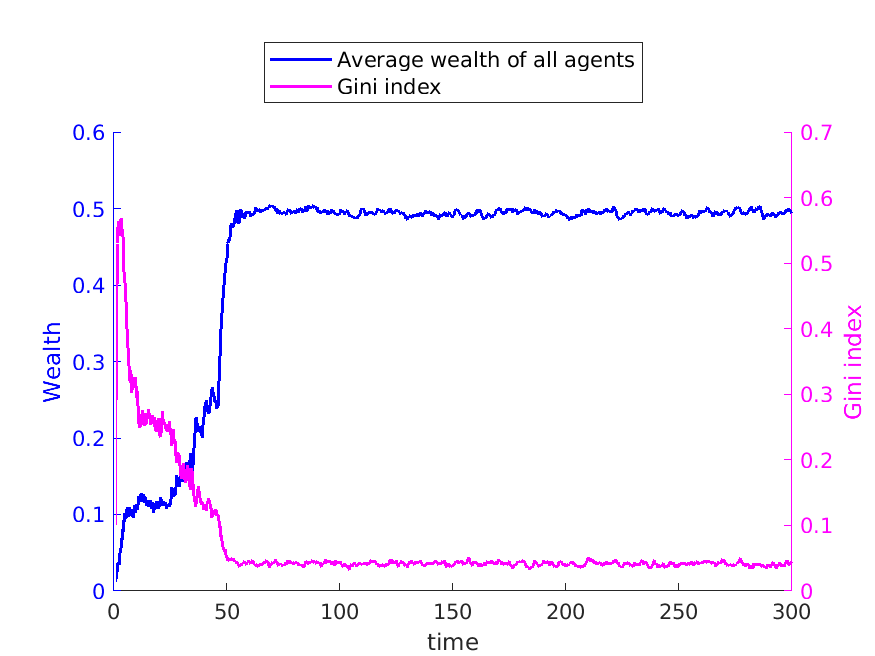}\label{fig9}}
\subfigure[$N_{coop}=20$, $N_{opp}=40$]{\includegraphics[width=0.49\textwidth]{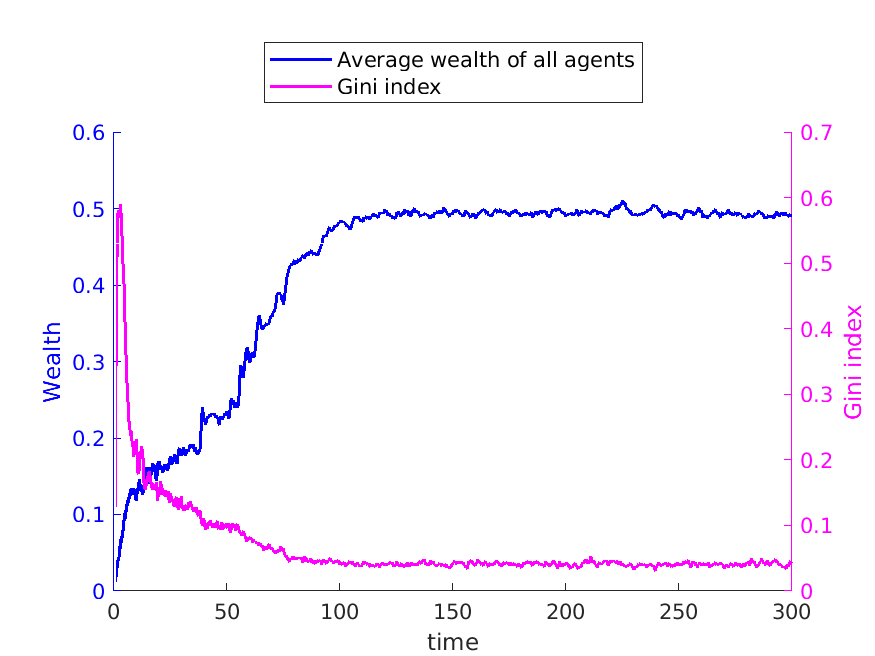}\label{fig10}}\\
\subfigure[$N_{coop}=30$, $N_{opp}=30$]{\includegraphics[width=0.49\textwidth]{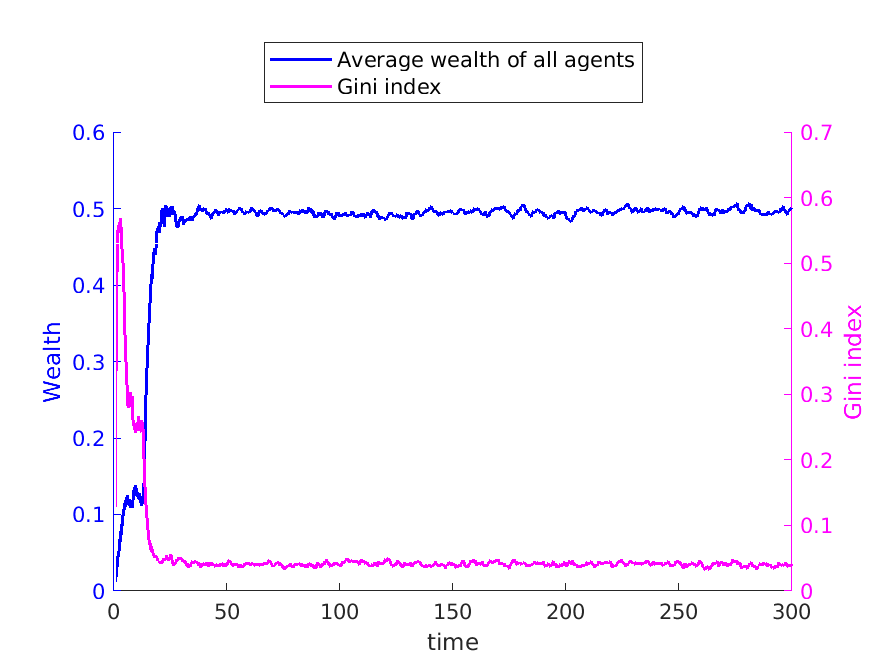}\label{fig11}}
\subfigure[$N_{coop}=40$, $N_{opp}=20$]{\includegraphics[width=0.49\textwidth]{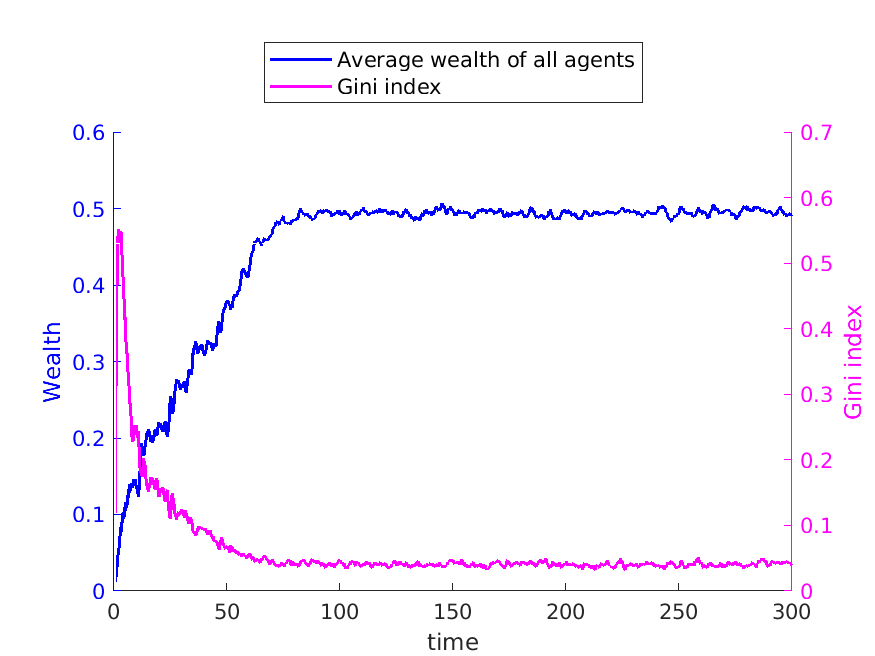}\label{fig12}}
\caption{$(\mathcal{H},\rho)$-induced dynamics with rules modifying both inertia and interaction parameters. Average wealth of all agents, and Gini index for the distribution of wealth among the agents of the system, with $N_{comp}=40$.  The scale on the left refers to wealth, that on the right to Gini index.}
\end{figure}

The results of the numerical simulations where both the inertia the interaction parameters are modified according to \eqref{rule_inertia} and \eqref{rule_interaction}, respectively, are displayed in Figures~\ref{fig9}-\ref{fig12}.
In these figures, we can no longer distinguish the initial subgroups as the rules modified the interaction parameters too, so that we plot the average wealth of all agents vs. time.

In this second scenario, the evolution of the network of agents is drastically changed. Even if the initial size of the cooperative subgroup seems to be not relevant (as in the first scenario), we observe that the average wealth   
of all agents has a rapidly increasing trend, and tends asymptotically to a value close to 0.6, much higher than the average wealth reached both using the classical Heisenberg dynamics and the $(\mathcal{H},\rho)$-induced dynamics where only inertia parameters are updated. Another interesting aspect to be underlined is concerned with Gini index. The initial value is 0 as we assigned to all agents the same amount of wealth, then, as time goes on, increase until a maximum value close to 0.5, then quickly decreases and approaches a very small value ($\approx 0.05$). This means that, although the average wealth status of the network is higher than that in the other scenarios, the system exhibits a very low level of inequality.

\section{Conclusions}
\label{sec:conclusions}

In this paper, we presented a fermionic operatorial model describing a system where some agents 
interact each other both with a competitive and/or a cooperative mechanism. The dynamics is assumed determined by a self-adjoint time-independent quadratic Hamiltonian operator. To enrich the dynamics, which using the classical Heisenberg view can be at most quasiperiodic, we adopted the approach of $(\mathcal{H},\rho)$-induced dynamics \cite{BDSGO-PhysicaA,fffbook}. 
Therefore, we superposed to the Heisenberg dynamics two different rules: the first one acts only on the inertia parameters, the second one both on inertia and interaction parameters. The resulting dynamics provided interesting results in this second scenario. In fact, the average wealth of all agents turned out to be  much higher, and the evolution was able to guarantee a very low degree of inequality.

What the simulations seem to show is that, whether we introduce the rules or not, 
cooperative agents tend to achieve an average wealth greater than that reached by competitive and opportunist agents. 

\section*{Acknowledgements}
Work carried out under the patronage of ``Gruppo Nazionale per la Fisica Matematica'' (GNFM) of the ``Istituto Nazionale di Alta Matematica'' (INdAM). M.G. and F.O. acknowledge financial support  by the PRIN grant \emph{Transport phenomena in low dimensional structures: models, simulations and theoretical aspects}, project code 2022TMW2PY, CUP B53D23009500006.

%\bibliographystyle{sageV}
%\bibliography{coop}
\end{document}